\documentclass[12pt]{elsart}

\usepackage{color,times}
\usepackage{amssymb,amsmath,graphicx,color,url}
\usepackage{lineno}
\usepackage{textcomp}
\usepackage[square,sort&compress,comma]{natbib}

\journal{Chemical Engineering Journal} 

\begin{document}

\begin{frontmatter}
\title{Long-term temporal dependence of droplets transiting through a fixed spatial point in gas-liquid two-phase turbulent jets}
\author[KL,ChE]{Li-Jun Ji},
\author[KL,SB,RCSE,SS]{Wei-Xing Zhou\corauthref{cor}},
\corauth[cor]{Corresponding author. Address: 130 Meilong Road, P.O.
Box 114, East China University of Science and Technology, Shanghai
200237, China, Phone: +86 21 64253634, Fax: +86 21 64253152.}
\ead{wxzhou@ecust.edu.cn} %
\author[KL,ICCT,RCSE]{Hai-Feng Liu},
\author[KL,ICCT,RCSE]{Xin Gong},
\author[KL,ICCT,RCSE]{Fu-Chen Wang},
\author[KL,ICCT,RCSE]{Zun-Hong Yu}
\address[KL]{Key Laboratory of Coal Gasification of Ministry of Education, East China University of Science and Technology, Shanghai 200237, China}
\address[ChE]{College of Chemical Engineering, East China University of Science and Technology, Shanghai 200237, China} %
\address[SB]{School of Business, East China University of Science and Technology, Shanghai 200237, China}
\address[RCSE]{Research Center of Systems Engineering, East China University of Science and Technology, Shanghai 200237, China} %
\address[SS]{School of Science, East China University of Science and Technology, Shanghai 200237, China}
\address[ICCT]{Institute of Clean Coal Technology, East China University of Science and Technology, Shanghai 200237, China}%

\begin{abstract}
We perform rescaled range analysis upon the signals measured by Dual
Particle Dynamical Analyzer in gas-liquid two-phase turbulent jets.
A novel rescaled range analysis is proposed to investigate these
unevenly sampled signals. The Hurst exponents of velocity and other
passive scalars in the bulk of spray are obtained to be 0.59$\pm
$0.02 and the fractal dimension is hence 1.41$\pm $ 0.02, which are
in remarkable agreement with and much more precise than previous
results. These scaling exponents are found to be independent of the
configuration and dimensions of the nozzle and the fluid flows.
Therefore, such type of systems form a universality class with
invariant scaling properties.
\end{abstract}


\end{frontmatter}

\typeout{SET RUN AUTHOR to \@runauthor}

\section{Introduction}

It is well known that \cite{Richardson-1922}'s picture of turbulent
cascade, in which large eddies break down into smaller ones, is a
multiplicative process. This hierarchical cascade in turbulence can
be described by fractal geometry \cite{Mandelbrot-1974-JFM}, which
is characterized quantitatively by the fractal dimension of
self-similar structure of turbulence
\cite{Mandelbrot-1983,Frisch-1996}. Both the experimental and
theoretical aspects have been studied in past decades.

Lovejoy investigated the fractal nature of satellite- and
radar-determined cloud and rain areas covering 6 orders of magnitude
of area sizes \cite{Lovejoy-1982-Science}. The area-perimeter
relation, introduced by Mandelbrot \cite{Mandelbrot-1983}, was found
to hold with the fractal dimension $D_{1}=1.35\pm 0.05$. A
theoretical analysis was proposed by Hentschel and Procaccia
\cite{Hentschel-Procaccia-1984-PRA}. They developed a theory of
turbulent diffusion and obtained the natural consequence that
$1.37<D_{1}<1.41$, which is in excellent agreement with the
empirical results of Lovejoy \cite{Lovejoy-1982-Science}.

Another important experiment concerns the studies on the fractal
facet of the turbulent-nonturbulent interface in turbulent flows by
Sreenivasan and his coworkers
\cite{Sreenivasan-Meneveau-1986-JFM,Sreenivasan-Ramshankar-Meneveau-1989-PRSA,Prasad-Sreenivasan-1990-PFA}.
Prasad and Sreenivasan applied the laser-induced fluorescence
technique to obtain the images of two-dimensional cuts of turbulent
jets \cite{Prasad-Sreenivasan-1989-EF,Prasad-Sreenivasan-1990-JFM}.
Applying the box-counting method, they estimated the fractal
dimension of turbulence interface in the K range and found that
$D_2=2.36\pm 0.05$ for all fully turbulent flows
\cite{Prasad-Sreenivasan-1990-PFA}. Taking into account the
influence of local fluctuations in the Kolmogorov scale on the
surface area due to the multifractal nature of the rate of
dissipation, slight revision was made
\cite{Meneveau-Sreenivasan-1990-PRA}. In addition, Huang {\em{et
al.}} proposed a modified box-counting method and found that
$D_{2}=2.43\pm 0.04$ in the K range for round jets
\cite{Huang-Li-Yu-1994-CSB}.

All the above experimental results are based on the box-counting
method, which may lead to several disadvantages and difficulties.
Difficulties associated with box-methods are typically attributed to
the lack of a significant scaling range, low signal-to-noise ratio
that limits reliable determination of level sets, and possible
inadequate records resulting in poor statistics confidence. Several
methods proposed to analyze box-counting statistics that do not
presume power-law behaviors have been employed to address the first
issue, which aim at determining the scaling range, if any, in an
unbiased fashion. In addition, the linear correlation coefficient in
log-log plots is not high enough, which results in relatively large
standard deviations. It is well know that, for fractals with
underlying cascade process, there are logarithmic periodic
oscillations
\cite{Badii-Politi-1984-PLA,Smith-Fournier-Spiegel-1986-PLA,Sornette-1998-PR,Johansen-Sornette-Hansen-2000-PD,Zhou-Sornette-2002-PD,Zhou-Sornette-Pisarenko-2003-IJMPC}.
The logarithmic periodic oscillations can be used to explain why
different samplings lead to different estimates of fractal
dimension. Zhou and Sornette proposed that canonical averaging of
variety of samplings can be used to eliminate logarithmic periodic
oscillations and get more reliable fractal dimension
\cite{Zhou-Sornette-2002-PD}.

Alternatively, rather than using the classic box-counting method,
Zhou {\em{et al.}} performed a rescaled range analysis upon signals
measured by Dual PDA to determine the fractal dimension of coaxial
turbulent jet, in which one-dimensional cuts are handled and the
unequally spaced time series were pre-processed using averaging to
form equidistantly spaced series \cite{Zhou-Wu-Zhao-Yu-2000-HGXB}.
The fractal dimension was found to be $D_1=1.42\pm 0.07$. Since the
rescaled range analysis is conducted based on averaging different
subseries at each scale, possible log-periodic oscillations are
eliminated. However, the moving averaging interpolation approach is
not satisfying. In this paper, we shall develop a variant of the
rescaled range analysis to make it suit directly for unequally
spaced time series such that the corresponding results are more
precise.

This paper is organized as follows. In Sec.~\ref{s1:RSA}, we propose
a generalization of the classical rescaled range analysis. Section
\ref{s1:result} reports the experiential results and corresponding
fractal dimensions for a variety of experiments under different
conditions. By comparing with interface dimensions of scalar field,
physical interpretation is presented in Sec. \ref{s1:diss}. Section
\ref{s1:conc} concludes.

\section{Rescaled range analysis}
\label{s1:RSA}

Rescaled range analysis, also termed as $R/S$ analysis or Hurst
analysis, was originally developed by Hurst \cite{Hurst-1951-TASCE}.
This analysis is based on a new statistical development and provides
an approach for analysis and characterization of time series which
has no underlying periodicity, yet retains long term correlation
\cite{Mandelbrot-Wallis-1969a-WRR,Mandelbrot-Wallis-1969b-WRR,Mandelbrot-Wallis-1969c-WRR,Mandelbrot-Wallis-1969d-WRR}.
The classic $R/S$ analysis is performed on a discrete time series
data set $\left\{ y_{i}:i=1,2,\cdots ,n\right\} $ with the time
$t_{i}$ uniformly-spaced, which is however not suitable for other
time series that are not equidistantly sampled, such as data
measured by Dual PDA. We thus generalize the classic $R/S$ analysis
from equidistant sampling to uneven sampling.

Define a continuous function $y:\left[ a,b\right] \rightarrow \Re $.
In a certain sense, graph
\begin{equation}
graph\left( y\right) =\left\{ \left( t,y\left( t\right) \right)
:t\in \left[ a,b\right] \right\}
\end{equation}%
may be regarded as a 2-dimensional fractal in the plane $\left(
t,y\right)$. Falconer has presented a method to estimate the fractal
dimension of functional graph \cite{Falconer-2003}. Let $\left[ \tau
_{1},\tau _{2}\right] \subset \left[ a,b\right] $. The time span
$s=\tau _{2}-\tau _{1}$ of $y$ is referred to as ``lag''
\cite{Mandelbrot-Wallis-1969b-WRR,Mandelbrot-Wallis-1969c-WRR}. Then
the mean of $y$ on $\left[ \tau _{1},\tau _{2}\right] $\ is
\begin{equation}
\left\langle y\right\rangle =\frac{1}{\tau _{2}-\tau
_{1}}\int\nolimits_{\tau _{1}}^{\tau _{2}}y\left( t\right) \rm{d}t.
\label{RS01}
\end{equation}
Define the cumulative deviation to be
\begin{equation}
X\left( t\right) =\int\nolimits_{\tau _{1}}^{t}\left[ y\left(
t\right) -\left\langle y\right\rangle \right] \rm{d}t \label{RS02}
\end{equation}
and the cumulative range to be
\begin{equation}
R=\underset{\tau _{2}\geqslant t\geqslant \tau _{1}}{\sup }X\left(
t\right) -\underset{\tau _{2}\geqslant t\geqslant \tau _{1}}{\inf
}X\left( t\right). \label{RS03}
\end{equation}
The standard deviation is
\begin{equation}
S=\left\{ \frac{1}{\tau _{2}-\tau _{1}}\int\nolimits_{\tau
_{1}}^{t}\left[ y\left( t\right) -\left\langle y\right\rangle
\right] ^{2}\rm{d}t\right\} ^{\frac{1}{2}}.  \label{RS04}
\end{equation}
Therefore, as a generalization of the $R/S$ analysis for discrete
time series, we assume that ${R}/{S}$ is self-affine and scales with
respect to $s$ as a power law
\begin{equation}
{R}/{S}\varpropto \left( \tau _{2}-\tau _{1}\right) ^{H}=s^{H},
\label{RS05}
\end{equation}%
where $H$ is termed as the Hurst exponent.

Partition the interval $\left[ \tau_1,\tau_2\right] $ into $n$
divisions with $\tau_1=t_0<t_1<\cdots <t_n=\tau_2$. Then
$y_{i}=y\left( t_{i}\right) $ is a unequally spaced time series.
Using rectangular approximation for integration, one can descretize
Eqs. (\ref{RS01})-(\ref{RS05}) as follows.The mean of $y$ can be
regarded as a time-weighted average
\begin{equation}
\left\langle y\right\rangle
=\frac{1}{t_{n}-t_{0}}\overset{n}{\underset{i=1}{\sum
}}y_{i}\times\left( t_{i}-t_{i-1}\right), \label{RS06}
\end{equation}
and the cumulative range becomes
\begin{equation}
R=\underset{1\leqslant j\leqslant n}{\max }\left\{ X_{j}\right\}
-\underset{1\leqslant j\leqslant n}{\min }\left\{ X_{j}\right\},
\label{RS08}
\end{equation}%
where
\begin{equation}
X_{j}=\overset{j}{\underset{i=1}{\sum }}\left( y_{i}-\left\langle
y\right\rangle \right) \left( t_{i}-t_{i-1}\right) \label{RS07}
\end{equation}%
is the cumulative deviation. Taking into account the weighted
standard deviation
\begin{equation}
S=\left\{ \frac{1}{t_{n}-t_{0}}\overset{n}{\underset{i=1}{\sum
}}\left( y_{i}-\left\langle y\right\rangle \right) ^{2}\left(
t_{i}-t_{i-1}\right) \right\} ^{\frac{1}{2}},  \label{RS09}
\end{equation}%
we have
\begin{equation}
{R}/{S}\varpropto \left( t_{n}-t_{0}\right) ^{H}=s^{H}. \label{RS10}
\end{equation}

If $t_{i}-t_{i-1}=\Delta t$ is a constant, the cumulative deviation
$X_{j}$ given in (\ref{RS07}) is $\Delta t$ times that in the
conventional $R/S$ analysis for uniformly spaced time series data.
However, it has no impact on the estimation of Hurst exponent using
Eq.~(\ref{RS10}). The Hurst exponent $H$ of the time series is then
evaluated from the slope of the straight line fitted to these
points. The estimated Hurst exponent is related to its fractal
dimension $D$ by the relation \cite{Mandelbrot-1983}
\begin{equation}
D=2-H.
\end{equation}
The $R/S$ analysis is a simple and robust method for fast fractal
estimate with as few as 30 data points
\cite{Handley-Jaenisch-Bjork-Richardson-Carruth-1993-Simulation}.

There are a variety of algorithms for performing $R/S$ analysis.
Consider a subseries $\left\{ y_{i},y_{i+1},\cdots
,y_{i+s-1}\right\} $, where $i$ is the departure of the sub-series
and $s$ is the length of the sub-series with $1\leqslant
i<i+s-1\leqslant n$. Different selection of $i$ and $s$ results in
different algorithm. In this paper, we propose to adopt a random
selection algorithm, in which $m$ starting points $i$ are chosen
randomly based on a uniform distribution on the interval
$[1,n-s+1]$. An average of the $m$ values of $R/S$ with the same lag
$s$ is calculated, which is denoted by ${R\left( s\right) }/{S\left(
s\right) }$. The Hurst exponent can be calculated according to
Eq.~(\ref{RS10}).

\section{Experimental measurement of fractal dimensions}
\label{s1:result}

\subsection{Experimental set-up}

The Dual PDA is based on a novel concept that yields a higher
measurement accuracy and performs non-intrusive measurements of the
velocity, diameter and transit time of spherical particles, droplets
and bubbles suspended in gaseous or liquid flows, particularly for
spray analysis and other investigations of liquid atomization. The
underlying principle of phase Doppler anemometry is based on
light-scattering interferometry and therefore requires no
calibration. The measurement point is defined by the intersection of
two or three pairs of focused laser beams and the measurements are
performed on single particles as they move through the measurement
volume. Particles thereby scatter light from the laser beams,
generating an optical interference pattern. A receiving optics
placed at a well-chosen off-axis location projects a portion of the
scattered light onto multiple detectors. Each detector converts the
optical signal into a Doppler burst with a frequency linearly
proportional to the particle velocity. The phase shift between the
Doppler signals from different detectors is a direct measure of the
particle diameter. We can obtain the arrival time and transit time
simultaneously.

\begin{figure}[htb]
\centering
\includegraphics[width=9cm]{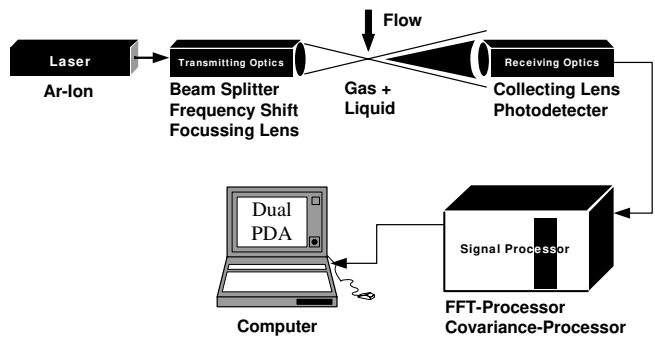}
\caption{Experimental set-up of the Dual PDA system}
\label{Fig:DualPDA}
\end{figure}

The Dual PDA receiving probe contains four receiving apertures
integrated into one single optical unit. The Dual PDA detector
configuration (2 standard and 2 planar), combined with sophisticated
validation routines, is not susceptible to unwanted effects
resulting from the Gaussian light intensity distribution in the
measurement volume. Misinterpreted size measurements due to
trajectory effects are therefore eliminated. The front optics module
for 3D-PDA configurations simplifies the alignment procedure
considerably. Screwed onto the PDA receiving probe and connected to
the transmitter optics by a dual-fibre link, the front optics module
generates the third pair of laser beams with adjustable beam
intersection and focus. The received signals are fed to one of
Dantec's advanced signal processors, which delivers results to a PC.
The instrument chart of the Dual PDA system is shown in
Fig.~\ref{Fig:DualPDA}. The power of the laser generator is 2W. The
focal length of the transmitting and receiving lenses are both
500mm.

\subsection{Experimental conditions}

In the experiments, we used tri-passage coaxial nozzles. The liquid
phase water moves through the mid-passage, while the gaseous phase,
say air, N$_{2} $ and CO$_{2}$, passes through the inner and outer
passages. In the experiments, the gaseous medium is changeable and
the fluid flow rates of the two phases are adjustable. The
measurement points distribute through out the spray zone. For a
fixed experimental configuration and given fluid flow, we record the
arrival time $t$, transit time $T$, axial velocity $U$, radial
velocity $V$ and drop diameter $d$ of 20000 drops moving through the
measurement point. In experiment (a), the flows of nitrogen in the
inner and outer passages are respectively 0.5 and 3.0m$^{3}$/h,
while those in the experiment (b) are $0.69$ and $3.9$m$^{3}$/h,
respectively.

For a fixed measurement point at $\left( 0,0,-26\right) $, the data
rate is $20.8$KHz, the mean velocity is $33.81$m/s, and the r.m.s.
velocity is $9.51$ m/s. The Kolmogorov microscale $\eta $ is
calculated from the signals according to
\begin{equation}
\eta =\left[ \frac{\nu ^{2}U_{1}^{2}}{15\left\langle \left( \partial
U/\partial t\right) ^{2}\right\rangle }\right] ^{1/4}~,
\end{equation}%
where $U_{1}$ is the mean speed at the measurement point, $\nu$ is
the kinematic viscosity of air, and $U$ is the streamwise component
of velocity. The resulting value of $\eta $ is $0.013$cm. The Taylor
microscale $\lambda
=0.41$cm is calculated according to%
\begin{equation}
\lambda =\frac{U^{\prime}U_{1}}{\left\langle \left( \partial
U/\partial t\right) ^{2}\right\rangle ^{1/2}}~,
\end{equation}%
where $U^{^{\prime }}$ is the r.m.s. of velocity fluctuations.
Hence, the Taylor-scale Reynolds number is
\begin{equation}
{\rm{Re}}_{\lambda }=U_{1}\lambda/\nu=273~,
\end{equation}%
which is moderate.

\begin{figure}[htb]
\centering
\includegraphics[width=7cm]{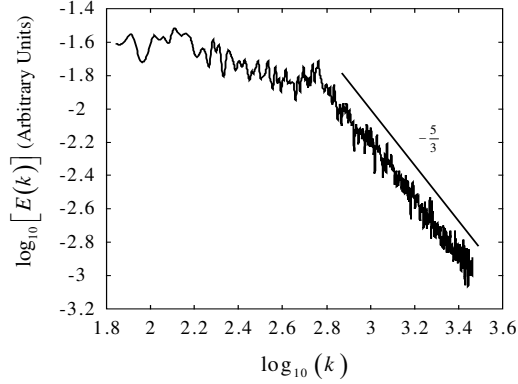}
\caption{Energy spectrum of velocity signals} \label{Fig:Sf}
\end{figure}

The power spectrum of the velocity signal is shown in
Fig.~\ref{Fig:Sf} which is obtained by the Dual PDA Processor using
the Gabor Fourier transform. A power-law scaling $E(k) \thicksim
k^{-\beta}$ with an exponent $\beta$ close to $5/3$ is observed over
a substantial range of about $0.7$ decades. Since the sampling
frequencies are not high enough, the highest frequency is in the
inertial range, which implies that we are dealing with the K-range
in this work.

\subsection{Hurst exponents and fractal dimensions}

\begin{figure}[htb]
\centering
\includegraphics[width=7cm]{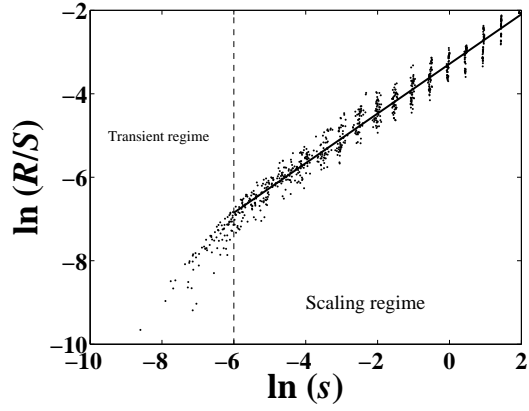}
\caption{A typical diagram of $\ln(R/S)$ versus $\ln(s)$}
\label{Fig:RSA}
\end{figure}

First of all, we analyzed the velocity signals arising from
experiments by random selection algorithm. A typical diagram of
$\ln(R/S)$ versus $\ln(s)$ is shown in Fig.~\ref{Fig:RSA}. There is
a transient regime for $\ln(s)<-6$, which exhibits a clear dropdown
compared with the scaling regime. The slope of the trend line in the
scaling regime gives the Hurst exponent $H=0.5926$. When adopting
the approximation that the measurement is evenly spaced, the Hurst
exponent estimated in the same scaling range is $\hat{H}=0.5518$,
which is significantly smaller than the real value of $H$. This
phenomenon is systematically observed for other experimental
realizations.

\begin{figure}[htb]
\centering
\includegraphics[width=7cm]{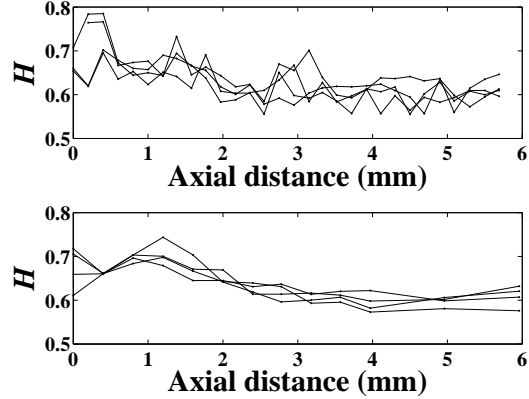}
\caption{Typical diagrams of Hurst exponent evolution along axial
direction} \label{Fig:H:Trans}
\end{figure}

According to the trend of Hurst exponents from the experiments, we
can roughly classify the turbulent spray zone into two parts: the
transient region and the steady region. In addition, the span of the
transient zone is found to be proportional to the spray speed. Two
typical diagrams of variation of Hurst exponent along the axial
direction are shown in Fig.~\ref{Fig:H:Trans}. It is obvious that
there exists transient region near the outlet of the nozzle. The
span of transient zone of (a) is about $2\sim 3$mm, while that of
(b) is about $3\sim 4$mm. One can see that $H$ decreases along the
streamwise direction near the outlet of the nozzle, which shows that
the signals close to the nozzle have stronger persistency. This is
induced by the stronger interactions among drops moving through the
measurement points. After a short transient period, the four types
of $H$ arrive a ``steady'' state, which fluctuates somewhat randomly
near the average $0.59$. We should point out that the measurement
range in the experiment with a distance from the outlet of the
nozzle of about $200$mm is much wider than what we have presented\
in Fig.~\ref{Fig:H:Trans}.

\begin{figure}[htb]
\centering
\includegraphics[width=7cm]{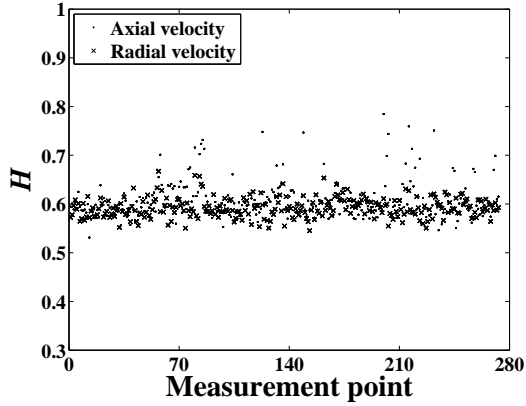}
\caption{Hurst exponents of velocity signals} \label{Fig:H:Velocity}
\end{figure}

The Hurst exponents of axial and radial velocities are plotted in
Fig.~\ref{Fig:H:Velocity}. The experiment conditions and/or the
spatial position of measurement are different to each other. The
hurst exponent is independent of experiment condition and the
spatial position. It is easy to find that, there are several points
with relatively higher $H$ of the axial velocity. These signals are
nonstationary and are not available for $R/S$ analysis, which will
be addressed later. Thus we withdraw these points.
It follows that%
\begin{equation}
H_{\rm{U}}=0.60\pm 0.02  \label{H_U}
\end{equation}%
and%
\begin{equation}
H_{\rm{V}}=0.59\pm 0.02.  \label{H_V}
\end{equation}%
In fact, the existence of nonstationary signals does not affect the mean
values, but they will increase the standard deviations.

\begin{figure}[htb]
\centering
\includegraphics[width=7cm]{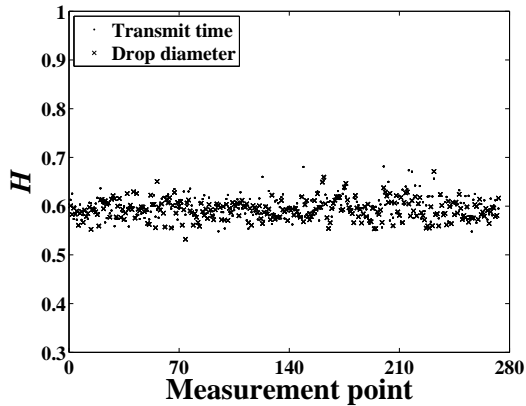}
\caption{Hurst exponents of transit time and drop diameter signals}
\label{Fig:H:Scale}
\end{figure}

Similarly, the Hurst exponents of transit times and drop diameters
are shown in Fig.~\ref{Fig:H:Scale}. Nonstationary signals,
especially of transit time, appear again. One can also find that,
when investigating a signal with high $H$ in Fig.\ref{Fig:H:Scale},
those signals from the same measurement record of the investigated
signal have high $H$ as well. Withdrawal of these nonstationary
signals follows that%
\begin{equation}
H_{\rm{T}}=0.60\pm 0.02  \label{H_T}
\end{equation}%
and%
\begin{equation}
H_{\rm{d}}=0.59\pm 0.02~.  \label{H_d}
\end{equation}

\begin{figure}[htb]
\centering
\includegraphics[width=7cm]{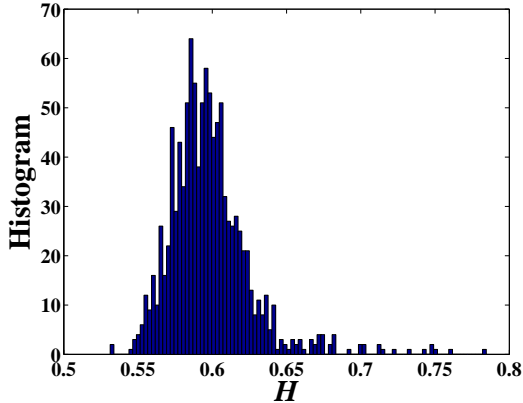}
\caption{Histogram of Hurst exponents of all signals}
\label{Fig:Hist}
\end{figure}

It is obvious that the Hurst exponents of different signals are
identical to each other. The difference among the four types of $H$
is within the scope of experiment error. Consequently, the
considered signals are also self-affine among each other. The
histogram of the Hurst exponent distribution of all signals is shown
in Fig.~\ref{Fig:Hist}. The majority of Hurst exponents concentrate
around the mean value $0.59$. The averaged Hurst exponent can thus
be calculated as
\begin{equation}
H=0.59\pm 0.02~.  \label{H_bulk}
\end{equation}%
We can hence obtain the fractal dimension of the signals that%
\begin{equation}
D=1.41\pm 0.02~.  \label{D_bulk}
\end{equation}

In a nutshell, the Hurst exponent and the corresponding fractal
dimension of the signals are independent of the spatial position of
the measurement position. That is, the investigated turbulent jet is
fractally homogenous in the main bulk of the spray zone. Moreover,
we can say that such two-phase flows form a universal class with a
universal Hurst exponent, since the fractal dimensions are invariant
with the changes of nozzle configuration, fluid medium, and flow
rates as well \cite{Zhou-Wu-Zhao-Yu-2000-HGXB}. However, we should
point out that the Hurst exponents change for other type of
two-phase flows. For example, coherent structure appears when
measuring signals of hydrogen jet into air
\cite{Liu-Wu-Wang-Gong-Yu-2000-HGXB,Liu-Zhao-Wang-Gong-Yu-2000-HGXB}.
Such a self-organized structure is expected to strengthen the
long-term dependence and decrease the fractal dimensions.

\section{Discussions}
\label{s1:diss}

\subsection{Nonstationary signals}

One may take it for granted that the computed Hurst exponent of a
fixed record is not reproducible when a repeated calculation is
carried out. However, we would like to point out the stability and
repeatability of the random sub-series algorithm, which has been
verified by repeating computations such that the resulting Hurst
exponents change very slightly compared with the previously
calculated values. Thus, it seems that the surprising high values of
Hurst exponents in Fig.~\ref{Fig:H:Velocity} and
Fig.~\ref{Fig:H:Scale} are unavoidable. Fortunately, we found that
signals corresponding to ``unexpected'' high values of $H$ are
nonstationary which can be used to distinguish these signals from
the rest. This is why we have got rid of experimental points with
high $H$.

A random process, say the axial velocity $U(t)$, is called
stationary if identical rules generate the process $U(t)$ itself and
all the processes deduced from $U(t)$ by a time shift, namely, all
the processes of the form $U(t+s)$. In experiments, we have to
control the conditions to be fixed through out the measurement of
one signal. Consequently, the mean of axial velocity component,
denoted as $\left\langle U\left( s\right) \right\rangle $ where the
averaging is performed over the time and $s$ is the sub-series
length, must be a constant. For nonstationary signals in the
experiments, an obvious change of $\left\langle U\left( s\right)
\right\rangle $ is detected in the plot of $\left\langle U\left(
s\right) \right\rangle $ against arrival time. In the present case,
the nonstationary signal can be regarded as a patchwork of several
stationary ones with different means. We find that, such a
scrambling leads to an upward in the tail of large lag $s$, namely,
the resulting Hurst exponent becomes higher. From a mathematical
point of view, $\left\langle \frac{R\left( t,s\right) }{S\left(
t,s\right) } \right\rangle $ for these nonstationary signals is
dependent of $t$ and thus $R/S$ analysis is not applicable
\cite{Mandelbrot-Wallis-1969d-WRR}.

\begin{figure}[htb]
\centering
\includegraphics[width=7cm]{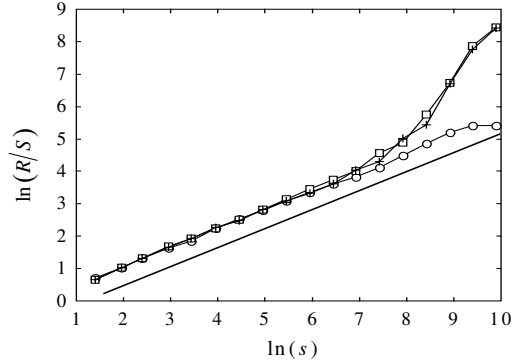}
\caption{Comparison of rescaled range analysis on the real time
series ($\circ$) and manipulated signals with $\Delta{u}=10$
($\small\square$) and $\Delta{u}=-10$ ($+$)}
\label{Fig:NonStationary}
\end{figure}

It is the usual situation that the flow flux decreases or increase
suddenly. Therefore, in order to verify the validity of the claim,
we performed numerical computations upon artificial series $u\left(
t_{i}\right) $ which
are generated using%
\begin{equation}
u\left(t_{i}\right) =\left\{
 \begin{array}{lccl}
 U\left( t_{i}\right),          &~~&{\rm{if}}&1\leqslant i\leqslant10000\\
 U\left( t_{i}\right) +\Delta u,&~~&{\rm{if}}&10001\leqslant i\leqslant20000
 \end{array}
\right.
\end{equation}%
from the real series $U\left( t_{i}\right) $,\ where $\Delta u$ is
an adjustable non-zero parameter. Denote $H^{a}$ for the Hurst
exponents of the artificial series. We find that $H^{a}\left( \Delta
u\right) $ is approximatively equal to $H^{a}\left( -\Delta u\right)
$ in a statistical sense, and $H^{a}$ increases with increasing
$\left| \Delta u\right| $ in the sampling time intervals for all
series. We use the same signal analyzed in Fig.~\ref{Fig:RSA} to
show the effect of non stead by comparing the results of raw signal
with manipulated signals ($\Delta{u}\ne 0$). For the sake of better
presentation, we treat all the signals as evenly spaced. The results
are shown in Fig.~\ref{Fig:NonStationary}. The circles represent the
real series, while the squares and pluses correspond to $\Delta
u=10$ and $\Delta u=-10 $, respectively. The solid line is used to
guide the eyes. We can see that the real series has scaling behavior
over the whole sampling interval. Nevertheless, the artificial
series have a more narrow scaling range and an upward end, which
results in the increase of the Hurst exponent.

Imagine a sudden drop in the flux of nitrogen in the outer passage
such that $\left\langle U(s) \right\rangle $ decreases. $
\left\langle T( s) \right\rangle $ will increase considerably due to
the strong dependence of transit time on the axial velocity, while $
\left\langle d(s) \right\rangle $ and $\left\langle V(s)
\right\rangle$ increases slightly. These changes result in the
in-step increases of $H$.

\subsection{Comparison with interface dimension}

It is natural to think that the passive scalars, the transit time
and drop size, are strongly affected by the turbulent velocity
field. Since the passive scalars are dependent of the turbulent
field, the fluctuation of passive scalars shows the situation of
turbulent field. If the investigated carriers are dense enough such
that they form a continuous turbulent-nonturbulent interface, we can
conjecture that the fractal dimension of the turbulent-nonturbulent
interface depend strongly upon the fluctuation of the velocity and
the passive scalars. We will find that the fractal dimensions of
interfaces and signals are essentially identical.

It is obvious that, if the drops in the spray zone is dense enough,
the spray can be regarded as a ``cloud''. Furthermore, if the
disperse phase in the spray aggregates to form larger bulk,
interfaces appear. Such crossovers illustrate the intrinsic
relationship among these systems. It is convenient to assume that
there are local imaginary interfaces in the investigated system. It
is well known that the area $S$ of the fractal interface can be
expressed as \cite{Mandelbrot-1983}
\begin{equation}
S\sim \eta ^{2-D_{2}}.
\end{equation}

We write $C$ for a scalar difference characteristic of large scales,
such as the r.m.s value, and $\eta $ for the Kolmogorov scale. It is
shown that the characteristic scalar gradient across interfaces is
of the order $C/\eta$
\cite{Sreenivasan-Ramshankar-Meneveau-1989-PRSA}. According to the
K41 theory, we have
\begin{equation}
\eta \sim \nu ^{3/4}.
\end{equation}%
Therefore, the flux of momentum across the interface can be written
as
\begin{equation}
{\rm{flux}}=\nu SC/\eta \sim \nu ^{(7-3D_{2})/4}. \label{flux}
\end{equation}
It is now well known that the growth rate of turbulent flows of a
given configuration are independent of fluid viscosity, which is
referred to as Reynolds number similarity
\cite{Prasad-Sreenivasan-1990-PFA}. It follows that
\begin{equation}
D_{2}=7/3.
\end{equation}

There are two possible corrections to $D_{2}$ according to
Eq.(\ref{flux}). By taking into account fluctuations in $\eta $ due
to the multifractality of the rate of energy dissipation, one may do
such correction by computing the mean value of ${C}/{\eta }$.
Further correction is made by considering the estimate of interface
area $S$ due to fluctuations in $\eta $. The corresponding results
is in essential agreement with the experimental outcomes.

Comparison of our results with the previously mentioned situations follows
that the fractal dimension $D$ of the signals is related to the interface
dimension $D_{2}$ by
\begin{equation}
D=D_{2}-1.
\end{equation}%
In other words, $D$ can characterize the turbulent-nonturbulent
interface as well as $D_{2}$. Actually, the signals fluctuate weaker
at the beginning of the spray and the interface, if exists, is much
smoother. This can account for the increasing trend of $D$ in the
transient region near the nozzle. In addition, the intrinsic
relevancy between the fluctuations of signals and the interface can
be also explained by Taylor's frozen flow hypothesis. Certainly, the
fluctuations of $D$ close to the mean is due to the variation of
mean velocity within a single record.

\section{Conclusions}
\label{s1:conc}

In this paper, we perform rescaled range analysis on the signals
measured by the Dual PDA in gas-liquid two-phase turbulent jets. We
generalize the classical R/S analysis to continuous form and then
discretize it to make it suitable for unequally spaced time series.
The fractal dimensions of the signals of axial and radial
velocities, transit time and drop diameter under different
experimental conditions are obtained both in the transient region
and the bulk of the spray zone. The fractal dimensions of signals of
different physical quantities are identical to each other.

In the transient region, since the liquid phase is being sped up by
the high-speed gaseous phase, the fluctuations become more and more
remarkable along the spray direction. As a consequence, the fractal
dimensions increase streamwisely. The length of the transient region
decreases with increasing gas flow rate. In the bulk of the spray
region, we have $D=1.41\pm 0.02$ for all signals and experiment
conditions, which is in excellent agreement with interface fractal
dimension by experiments and theoretical derivations and is not
effected by the drop size distribution. Since the fractal dimension
is invariant for experimental configurations and conditions, the
gas-liquid two-phase turbulent jets investigated in this work form a
universality class with an invariant exponent.

Certainly, there are still open problems. To address why the Hurst
exponents (and corresponding fractal dimensions) of different
variables at different measurement points are identical, we have to
uncover the underlying physics. However, we can only provide a
qualitative explanation rather than quantitative expressions. But to
make further progress, we would need some kind of theory or model to
give the power law relationship between the different variables. In
addition, it is necessary and interesting to test the log-periodic
oscillations in the turbulent signals, which may provide an
important step towards a direct demonstration of the Kolmogorov
cascade or at least its hierarchical imprint
\cite{Zhou-Sornette-2002-PD,Zhou-Sornette-Pisarenko-2003-IJMPC}.

\bigskip
{\textbf{Acknowledgments:}}

This work was partially supported by the National Basic Research
Program of China (No. 2004CB217703), the PCSIRT (IRT0620), the
Program for New Century Excellent Talents in University
(NCET-05-0413 and NCET-07-0288), and the Project Sponsored by the
Scientific Research Foundation for the Returned Overseas Chinese
Scholars, State Education Ministry.

\bibliographystyle{elsart-num}
\bibliography{E:/Papers/Auxiliary/Bibliography_FullJournal}

\end{document}